\documentclass[agupp,epsfig,graphicx]{aguplus}
\usepackage[dvipsone]{epsfig}
\authoraddr{W.~A.~Traub and K.~W.~Jucks, 
Harvard-Smithsonian Center for Astrophysics, 
60 Garden Street, Cambridge, MA 02138.\\
(email: wtraub@cfa.harvard.edu, kjucks@cfa.harvard.edu)} 

\begin{document}
 
\title{A Possible Aeronomy of Extrasolar Terrestrial Planets}
 
\author{W.~A.~Traub and K.~W.~Jucks}
\affil{Harvard-Smithsonian Center for Astrophysics, Cambridge, Massachusetts}
 
\begin{abstract}

Terrestrial planetary systems may exist around nearby stars as the
Earth-sized counterparts to the many giant planets already discovered 
within the solar neighborhood.  In this chapter we first discuss the 
numerous techniques which have been suggested to search for
extrasolar terrestrial planets.  We then focus on the expected results from
that technique in which an orbiting telescope or interferometer is used 
to obtain a visible or infrared spectrum of a planet, without 
contamination from the parent star.  We show examples of such spectra 
for selected cases: the present Earth, the Neoproterozoic (snowball) 
Earth, a methane-rich Earth, and the present Mars and Venus.  We conclude 
by discussing the implications of such spectra for the detection of life 
on an extrasolar terrestrial planet.

\medskip

{\it
To appear in ``Atmospheres in the Solar System: Comparative Aeronomy'', 
edited by M. Mendilo, A. Nagy, and H.J. Waite, 
AGU Geophysical Monograph 130, 369-380, 2002.}

\end{abstract}


\begin{article}
 
\section{KNOWN EXTRASOLAR SYSTEMS} 

The first planet orbiting a solar-type star beyond the solar system 
was announced in 1995.  Since then, as of October 2001, 66 planets 
have been found and confirmed, orbiting 58 stars with a median distance
from the sun of about 28 pc.  The search database contains roughly 
1200 stars.  

Current estimates of the frequency of massive planets range from
about $3$-$5$\% to $6$-$7$\% 
[{\it J.~Schneider, and G. Marcy, resp., personal communication}].  
The number of detected planets is growing
monthly, as observing techniques are refined, and as the time base of 
the record increases, allowing longer-period planet signatures to be 
extracted from the radial velocity sequences.

The definition of a planet can be a controversial issue, as is
evidenced by the recent debate on whether Pluto should be classified 
as a planet, or a trans-Neptunian or Kuiper-belt object.  
However in the case of extrasolar planets
the debate centers not on the question of the low-mass end of the
scale, as for Pluto, but on the high-mass end, which is the
realm of the planets discovered to date.  The question is, should these
objects be classified as planets, brown dwarfs, or small stars? 

Here we adopt the recommendation of {\it Oppenheimer et al.} [2000] 
for objects of solar metallicity:  the minimum mass for a brown dwarf 
is about 13~M$_J$ (where M$_J$ is the mass of Jupiter) 
or 0.013~M$_{sun}$, sufficient to allow deuterium 
burning;  the minimum mass for a main sequence star is about 78~M$_J$ or 
0.075~M$_{sun}$, sufficient to allow hydrogen burning.

Pulsar planets, known since 1992 from variations in pulse arrival 
times from pulsars, are kept in a separate category from exoplanets 
around main-sequence stars.  These bodies may have formed during the 
explosion that created the neutron-star pulsar.  

At least two websites collect current information on exoplanets.
One site is ``The Search for Extrasolar Planets'' [{\it Marcy, 2001}]
originating at UC Berkeley.
The other, larger, site is ``The Extrasolar Planets Encyclopaedia'' 
[{\it Schneider}, 2001] originating at Paris Observatory, Meudon.
Both sites feature news items, discussions, tutorials, 
papers, bibliographies, and comprehensive lists of exoplanets, their 
properties, and their parent stars.  Both sites are expertly edited
and authoritative.

The present paper is oriented toward extrasolar terrestrial planets, 
i.e., those in the mass and temperature range of Venus, Earth, and Mars. 
However, since all known exoplanets are in the gas giant mass range, 
it is appropriate to ask about the prospects for finding 
terrestrial-mass exoplanets.  

Astrometry (cf. later section) tells us the semi-major axis, eccentricity, 
and mass of a planet.  Each of these is a clue to a part of the history of 
the planet, e.g., where and how it was formed, whether it has migrated
since then, and whether its orbit has been perturbed by other bodies. 
However to learn from this data if terrestrial-sized planets might be 
present we must fall back on inference based on the observed
frequency distribution of mass, as follows. 

If we examine a plot of exoplanet mass versus semi-major axis, we see that 
the discoveries tend to populate the entire region between the extremes of: 
(a) maximum mass set by definition at about 13 $M_J$; 
(b) minimum mass set by the radial velocity detection method 
at about $M/M_J = 0.035 v R^{1/2}$ where $v$ (m/s) is the minimum detectable 
orbital velocity amplitude and $R$ (AU) is the orbital radius;
(c) maximum radius set by a total observing time of about 5 years, 
to get a full orbit; and 
(d) minimum radius set by the smallest observed orbit at about 0.01 AU.
If we assume that this entire region, on a $(\log R, \log M)$ plot,
might be uniformly populated, then we can calculate the expected number
of planets as a function of mass.  

\begin{figure}[tb]
\figurewidth{8.4cm}
\begin{center} 
\epsfig{file=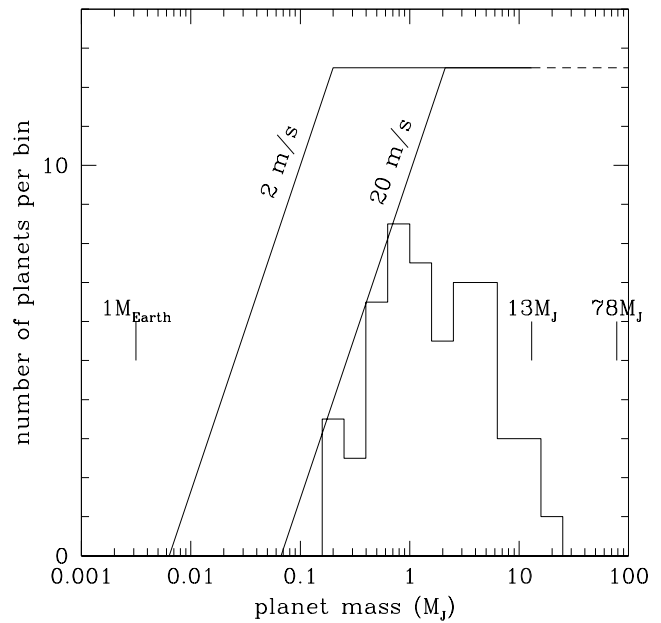, bbllx=20, bblly=174, bburx=576, bbury=708,
width=8.4cm}
\end{center} 
\caption{Histogram of discovered exoplanets, ranging from about 0.16 to
13 M$_J$.  Theoretical curves are shown for
the cases where exoplanets have a uniform density distribution in the
($\log M, \log R$) plane, in the range $\log M = -3$ to $+1.11$, 
and $\log R = -2$ to $+1$, and limited by the sensitivity of radial
velocity measurements to orbital velocities of 2 and 20 m/s, 
as indicated.  The observed distribution appears to be bounded by 
measurement accuracy on the low-mass side, and by a lack of exoplanets 
on the high-mass side.}
\label{fig0} 
\end{figure}

We show in Figure~\ref{fig0} a histogram of the detected masses, 
and two ``expected'' distributions corresponding to velocity amplitudes 
of 2 and 20 m/s, the former being a nominal goal of the present radial
velocity searches, and the latter being an estimate of the current
level of confident detection.  Although  Figure~\ref{fig0} is
undoubtedly overly simple, we may nevertheless draw two conclusions. 
(1) At the high-mass end, where the present searches are certainly
relatively complete and unbiased, the observed distribution drops off
rapidly from about 1 M$_J$ to 13 M$_J$, suggesting that the
exoplanet population really does exist as a separate entity from
any brown dwarf population (not shown) which might fill in the
13-75 M$_J$ range.
(2) At the low-mass end, the observed population seems to be
approximately fit by the 20 m/s curve, suggesting that the observations
are observationally limited, and that there is no evidence for a
fall-off at low masses.  Thus at present we may well be sampling only
the very top end of the exoplanet mass distribution curve, and it may
well be that the distribution function continues all the way down to
the M$_{Earth}$ range.   On this basis we optimistically look forward 
to someday finding terrestrial exoplanets.

\section{DETECTION METHODS: GRAVITY AND AERONOMY} 

There is a surprising number of proposed techniques to detect or 
characterize exoplanets.  About one-half of the techniques may be 
classified as being gravitational in nature, and the other half as being 
aeronomic (or photonic).
The gravitational methods are 
(1) radial velocity,
(2) astrometry,
(3) transits,
(4) pulsar timing,
(5) gravitational lensing, and
(6) disk shaping.
The aeronomic methods are
(7) visible-infrared shift,
(8) reflected light,
(9) transmitted light,
(10) auroral emission, 
(11) radio emission, 
(12) anthropogenic transmission, 
(13) coronagraphic imaging, and
(14) interferometric imaging.
 
\subsection{Radial velocity}  
As the planet and star orbit their common
center of mass, the velocity vector of the star projected along the
observer's line of sight is proportional to $M \sin i$
where $M$ is the planet mass and $i$ is the inclination of the orbit
plane to that of the sky.  Doppler shifts in stellar spectra
have been measured to an accuracy of about 3 m/s;  10 m/s is common,  
and 1 m/s may be the ultimate limit of this technique
[{\it Marcy and Butler}, 2000; {\it Santos et al.}, 2000].
By comparison, the solar velocity due to Jupiter is about 3 m/s, 
and that due to Earth is about 0.01 m/s.
 
\subsection{Astrometry}  The projection of stellar orbital motion onto 
the plane of the sky produces an astrometric shift which is measured
with respect to a grid of nearby reference stars.  
At 10 pc, Jupiter would move the Sun by
about 100 $\mu$as (1 $\mu$as $ = 10^{-6}$ arc-sec), and Earth would move
the Sun by about 0.3 $\mu$as.  The Hipparcos satellite had an accuracy
of about 500 $\mu$as, so could not quite detect Jupiter-sized planets,
but planned missions such as 
FAME (50 $\mu$as) [{\it Horner et al.}, 2001] (2004 launch planned),
GAIA (2-10 $\mu$as) [{\it Perryman et al.}, 2001], (2012 launch planned),
and SIM (1 $\mu$as) [{\it Danner and Unwin}, 1999] (2009 launch planned),
have a good chance of detecting masses in the sub-Jupiter range, 
and almost down to the terrestrial limit.

\subsection{Transits}  If the planet's orbital plane is seen nearly
edge-on, a partial eclipse of the star by the planet may occur.  For
the one example known to date, HD209458,  [{\it Charbonneau et al.}, 2000], 
precise photometry has allowed us to infer the stellar limb
darkening, the planet radius, the orbital inclination, and therefore
the planet's mass.  Further observation may lead to a transmitted
light measurement, as discussed below.  
Dedicated, staring telescopic searches [{\it e.g., Borucki et al.}, 2001] 
may detect more examples of this rare transit phenomenon.

\subsection{Pulsar Timing}  The clock-like constancy of pulsar spin rates 
means that 
the time delay produced by line of sight displacement due to an orbital
companion can be interpreted in terms of the mass and orbit of the
companion [{\it e.g., Konacki et al.}, 2000].  
Two such pulsars are known, PSR 1257+12 (3 planets, possibly 4), and 
PSR B1620-26 (1 planet).   Interestingly the 
technique is sufficiently sensitive that we know that some of these
pulsar planets are in the few $M_{Earth}$ range, and one may be in the 
$M_{Pluto}$ range.

\subsection{Gravitational Lensing}  A star in our Galaxy can cause a 
distant light source (a background galaxy) to apparently brighten for
several days by gravitationally deflecting and lensing the distant
galaxy's light as the star happens to pass in front of the galaxy.   
If the star hosts a planetary companion, then the planet can cause a 
secondary brightening [{\it Gaudi and Gould}, 1997].  The effect would be 
transient, but statistics of many events would give information on the 
Galactic incidence of planets.  

\subsection{Disk Shaping}  For a star with a debris disk, the presence of a
planet with non-zero eccentricity or relative orbit inclination will
cause the disk to become eccentric or warped, respectively;
a planet inside the dust or debris disk can also generate
resonant trapping (clumps) and clearing of central holes 
[{\it Wyatt et al.}, 1999; {\it Liou and Zook}, 1999].

\subsection{Visible-Infrared Shift}  Precise astrometry of an
unresolved star-planet system in the visible and infrared wavelength 
regions simultaneously would show a spatial shift because the star light 
dominates the planet's light less in the infrared.  This is a detection 
mode planned for the Keck Interferometer and Very Large Telescope 
Interferometer [{\it Akeson et al.}, 2000; {\it Lopez et al.}, 2000].

\subsection{Reflected Light}  For a large, close-in planet, sufficient
star light may be reflected from the planet that it could be directly detected as an
additional component of star light, even if the system is not
spatially resolved [{\it Charbonneau et al.}, 1999].

\subsection{Transmitted Light}  If a planet eclipses the parent star,
then the planet's transparent upper atmosphere will transmit a portion 
of the starlight, with a superposed planetary absorption
spectrum which could produce measurable features such as He 1.0830
$\mu$m or Na 0.589 $\mu$m [{\it Brown et al.}, 2001].

\subsection{Auroral Emission}  Auroral activity such as seen on Earth or
Jupiter (e.g., O 0.5577 $\mu$m, H$\alpha$) is a non-thermal, 
potentially useful indicator of a planet [{\it e.g., Waite et al.}, 2001],
but the flux rate may be so low as to not be competitive with other 
techniques.  Detection of oxygen emission is discussed more fully in
Chapter VI.2.

\subsection{Radio Emission}  Decametric radio wavelength radiation from
electrons in the magnetic field around Jupiter and
Io suggests that planets might be identified by this
non-thermal radio signature [{\it Bastian et al.}, 2000], 
but low flux might limit this method.

\subsection{Anthropogenic Transmissions} 
(a) It has been pointed out that a directed visible laser beam from Earth 
could be made to outshine the sun for the duration of a pulse, and would 
therefore be easily visible with a modest telescope at interstellar 
distances [{\it Howard and Horowitz}, 2001].  In the hope of detecting such 
pulses trial experiments have been started.  
(b) Pulsed radio transmissions containing coded messages are likewise
detectable at great distance, and have been the basis of several
searches [{\it e.g., Horowitz and Sagan}, 1993; {\it Leigh and Horowitz}, 1999].

\subsection{Coronagraphic Imaging}  
Direct visible-wavelength detection of the analog of
the solar system's gas giant outer planets could be achieved using
existing general-purpose telescopes, such as the Hubble Space Telescope 
or the planned Next Generation Space Telescope, if the residual optical 
imperfections in either telescope were to be corrected by adaptive optics. 
Dedicated coronagraphic telescopes with shaped or shaded pupils have also 
been proposed [{\it Nisenson and Papaliolios}, 2001; {\it Spergel}, 2001]. 
To detect an Earth at 10 pc and 0.5 $\mu$m wavelength
requires $10^{-10}$ starlight rejection at 0.1 arcsec separation.

\subsection{Interferometric Imaging}
An infrared-wavelength imaging interferometer has been proposed as a NASA
mission [{\it Beichman et al.}, 1999], or European Darwin mission 
[{\it Fridlund}, 2000].  To detect an Earth at 10 pc and 10 $\mu$m wavelength
requires $10^{-7}$ starlight rejection at 0.1 arcsec separation.

\section{WHY SPECTRA OF TERRESTRIAL PLANETS?}

Of all the techniques mentioned to detect or characterize an extrasolar 
terrestrial planet, we have chosen to focus on just two: coronagraphic 
imaging and interferometric imaging.  The reason for this is that these 
seem to give us the best chance to determine the atmospheric constituents 
of the planet, by direct observation of the reflected or emitted light from
the atmosphere and surface.

The current thumbnail picture of planetary formation is as follows.
A massive
molecular cloud is somehow triggered to collapse; a star is formed; the
star is surrounded by a remnant gas and dust cloud; rocky, metallic, and,
in the outer, colder parts of the cloud, icy grains, condense and 
agglomerate in the surrounding cloud; and the agglomerations cascade to 
larger sizes to form planetary cores.  Then in the outer part of the cloud,
where plenty of gas is available, the gas continues to collapse around the
cores, and Jupiter-like (gas giant) planets form.  In the inner part of the
cloud  where it is hotter and less gas is available, only rocky
planets form.  These rocky, terrestrial-type planets can have abundant 
liquid water or ice, and relatively thin atmospheric envelopes, both
generated by outgassing of the rocky material and possibly also by 
infalling comets.  It is on these planets, with their solid surfaces where 
water can accumulate and chemical reactions occur, and their thin
atmospheres where sunlight can penetrate and be used for driving chemical
reactions, that we speculate that life probably has its best chance to 
develop. 
(Excellent references can be found in 
{\it Chyba et al.} [2000],
{\it Lunine} [1999], and
{\it Yung and DeMore} [1999]).  

Once we know the abundances of key gases, we can then make informed
speculations on the likelihood that life exists on the planet.
However even with the restriction to terrestrial-type planets, we still
have a large range of possible types of atmospheres.  In the following
sections, we begin to explore the range of possibilities.  We start with a
description of our method for calculating visible and infrared spectra for
the case of the present Earth.  We extend this to include 
Neoproterozoic icehouse and hothouse Earths, a methane-rich Earth, and
the present Mars and Venus.  We conclude with a brief discussion
of how the presence of life might be inferred from spectra such as these.

\section{PLANETARY SPECTRA: PRESENT EARTH} 

\begin{figure}[tb]
\begin{center} 
\leavevmode 
\epsfxsize=3.25in
\epsffile[20 174 576 708]{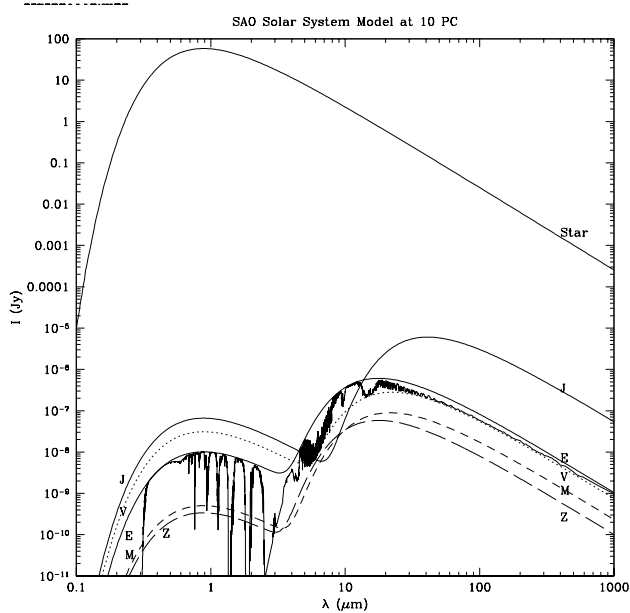}
\end{center} 
\caption{Solar system blackbody thermal emission spectra and reflected light 
spectra at 10 pc, for the Sun, Jupiter (J), Earth (E), Venus (V), Mars (M), and 
zodiacal dust (Z). For curve Z a telescopic field of view of 0.010 arcsec 
diameter centered at 0.1 arcsec from the star is assumed.    
For the Earth, present atmospheric abundances are used to calculate a line 
by line spectrum for the entire 4 decades in wavelength, for a cloud-free 
atmosphere.  (Note 1 Jansky is $10^{-26}$ watt m$^{-2}$ Hz$^{-1}$.)}
\label{fig1} 
\end{figure}

A graphical overview of the exoplanet detection issue is shown in
Figure~\ref{fig1}, 
where we plot the flux density of a model solar system
as it would be seen from a distance of 10 pc, the median distance 
for the nearest 450 or so stars in our Galaxy.   

Blackbody spectra of the Sun, Jupiter, and the 3 terrestrial planets 
are shown in Figure~\ref{fig1}, for the effective temperatures
of these bodies, except as follows:
the spectrum of Jupiter includes its internal heat source contribution;
the spectrum of Mars is an average of the day and night side spectra
at different temperatures; and for Earth the average ground
temperature is used, not the effective temperature. 
(N.B., the effective temperature of an object is the temperature 
of a blackbody which has the same area and total radiated
power as does the object.)

The zodiacal dust cloud is modeled as a face-on, smooth, optically thin
blackbody emitter with optical depth varying as $r^{-0.39}$ and temperature 
varying as $r^{-0.42}$ where $r$ is distance from the central star 
[{\it Reach}, 1995].

The reflected light from Jupiter
and the terrestrial planets is approximated by a scaled version of the
solar spectrum, proportional to each planet's average visual albedo
and area, reduced by a factor of 0.26 to approximate the brightness at
quadrature, when the observer sees only one-half of the disk
illuminated.  The zodiacal dust reflection spectrum is modeled with the 
same density distribution as for the thermal emission but with an albedo
selected to agree with visual observations. 

The spectral line component of the Earth's spectrum in 
Figure~\ref{fig1} is
calculated separately for the thermal emission and reflection cases,
for a clear atmosphere, and the results combined, as described next.

\subsection{Spectral Computation Method}

Model Earth spectra are calculated with our SAO code originally developed to
analyze balloon-borne far-infrared thermal emission spectra of the stratosphere
[{\it e.g., Traub and Stier}, 1976;  {\it Johnson et al.}, 1995],
extended to include visible reflection spectra.  The spectral line
data base includes the large AFGL compilation [{\it Rothman et al.}, 1998] 
plus improvements from pre-release
AFGL material and our own sources.  In a few cases laboratory cross
section spectra are available
but spectroscopic analysis is not, so here we use an empirical pseudo-line band
shape.  The far wings of pressure-broadened lines can be non-Lorentzian at
around 1000 times the line width and beyond, so in some cases 
(H$_2$O, CO$_2$, N$_2$) we replace the far wings of the line-by-line 
calculation with measured continua data in these regions.
Dust and Rayleigh scattering are approximated by empirical wavelength power laws
and contribute significantly only in the visible blue range.  
Model atmospheres from 0 to
100 km altitude are constructed from standard models discretised to appropriate
layers, and additional radiative transfer methods used to ensure that line cores
and optically thick layers are accurately represented.  

Radiative transfer from layer to layer is explicitly calculated using the 
average absorption and emission properties of each layer; scattering as a
source is neglected.  Integration from the spherical Earth atmosphere is 
approximated to a few percent accuracy by a single-point calculation at a 
zenith angle of 60 degrees, so the effective air mass is 2 in the infrared 
(outgoing emission) and 4 in the visible (2 for incoming sunlight, plus 2 
for outgoing reflected light).  

Cloud effects are beyond the scope of this chapter, and they are not included in the
calculations shown here, but they can be represented by inserting
continuum absorbing/emitting layers at appropriate altitudes;  broken clouds
can also be represented by a weighted sum of spectra using different cloud layers.
In general, the effect of clouds is to dilute the strength of line features in the
visible, and to dilute, but in extreme cases cause absorption lines to appear as
emission lines, in the infrared.

\subsection{Thermal Emission Spectrum}

\begin{figure}[tb]
\begin{center} 
\leavevmode 
\epsfxsize=3.25in
\epsffile[52 120 558 680]{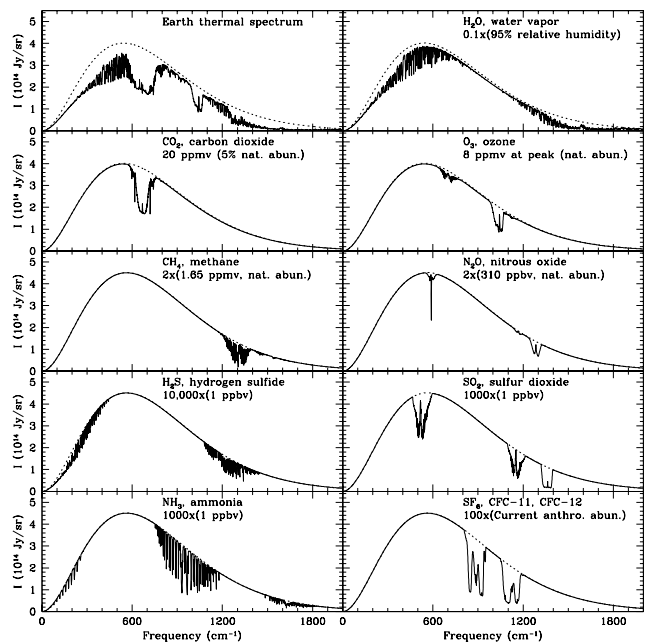}
\end{center} 
\caption{A calculated thermal emission spectrum of the present Earth
is shown in the top left panel.  The other panels show spectra of individual 
species, with their mixing ratio profiles scaled up or down so as to generate 
a maximum feature depth of roughly one-half.
The calculations are performed at very high resolution and subsequently 
smoothed to 1 cm$^{-1}$ for display.  
The panels demonstrate the concentrations needed for each
species to contribute significant opacities on an Earth-like planet.}
\label{fig2} 
\end{figure}

The dominant features of the Earth's thermal emission spectrum are illustrated 
in Figure~\ref{fig2}, 
where the blackbody flux and composite spectrum (Jy/sr) are shown in the
top left panel, and the other panels show the relative intensities of the 
major infrared molecular species (H$_2$O, O$_3$, CH$_4$, CO$_2$, N$_2$O) 
as well as minor contributors (H$_2$S, SO$_2$, NH$_3$, SF$_6$, CFC-11, CFC-12).  
The composite spectrum is calculated for the
present abundances of each species, but the individual species spectra are
calculated for increased or decreased abundances, with the expected vertical
mixing ratio profiles scaled so as to show the
absorption spectrum minima at an optical depth of about unity.

The H$_2$O panel shows the far-infrared rotational band and the mid-infrared
vibrational band, calculated for an abundance of about 0.1 times the saturated 
value in the lower troposphere. 
Even at this reduced concentration, the water lines are quite strong, but also rather
diffuse, without any well-defined compact spectral features in the thermal
infrared.  

Carbon dioxide, on the other hand, even at 5\% of present abundance, shows
a very strong 15 $\mu$m (667 cm$^{-1}$) band in the infrared, the depth of 
which is limited not by abundance, but rather by the thermal structure of 
the Earth's atmosphere, such that the minimum brightness corresponds to the 
blackbody strength at the altitude at which the band core optical depth 
reaches values on the order of unity.  Note the small spike in the center 
which is generated at an altitude of about 30 km by the temperature 
inversion in the stratosphere, causing the very strong core to appear in 
emission against the lower-lying cooler layers from which the near-wing 
emission emanates at an altitude of about 20 km. 

Ozone is the third most prominent infrared feature with its strong 
9 $\mu$m (1100 cm$^{-1}$) band, shown here at its natural abundance.  
The ozone feature is almost entirely due to the stratospheric O$_3$ layer, 
though there is a trace of tropospheric O$_3$.

Methane and nitrous oxide are both shown at twice natural abundance, and have
significant features nearly overlapping in the 7 $\mu$m (1400 cm$^{-1}$) region, 
also lying in the red wing of the 6 $\mu$m (1600 cm$^{-1}$)water band, and 
therefore not readily separable, but nevertheless in principle measureable.  
The combined effect of the current abundance of CH$_4$ and N$_2$O is seen 
in the composite panel, where the spectrum shows a rather sharp decrease 
going from about 1200 to 1300 cm$^{-1}$, compared to the gradual decrease 
in this region due to H$_2$O alone, as shown in the water-only panel. 

The remaining species, at present Earth abundances, are not expected to
be easily detectable on an exoplanet, due to the weakness of their bands.  The
panels show H$_2$S at 10,000 times natural abundance, SO$_2$ at 1000 times, 
NH$_3$ at 1000 times, and the anthropogenic gases SF$_6$, CFC-11, 
and CFC-12 each at 100 times current abundance.

\subsection{Reflection Spectrum}

\begin{figure}[tb]
\begin{center} 
\leavevmode 
\epsfxsize=3.25in
\epsffile[50 140 527 635]{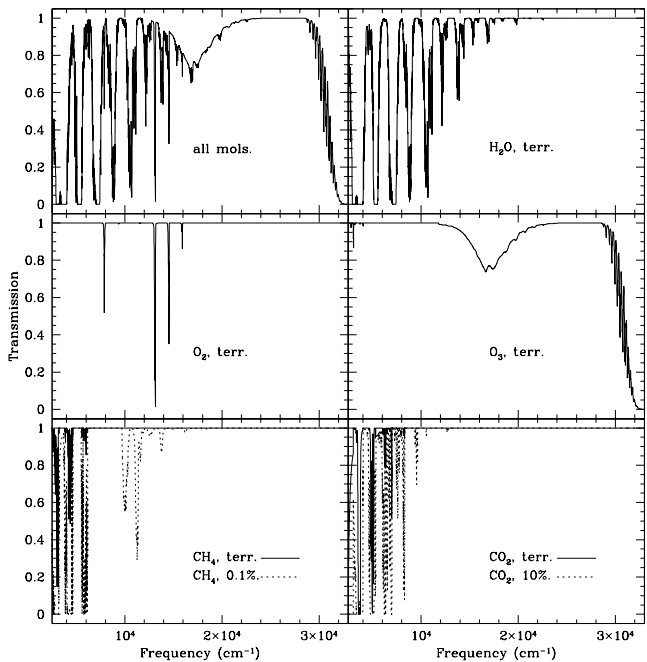}
\end{center} 
\caption{The reflectivity of the Earth with present atmospheric abundances 
is shown in the top left panel, normalized to unity.  Clouds, aerosols, and
Rayleigh scattering are ignored in this example.  The other panels show 
reflection spectra for the cases where only a single species is present in
terrestrial abundances.  For CH$_4$ and CO$_2$ we also show spectra for greatly
enhanced abundances as discussed in the text.}
\label{fig3} 
\end{figure}

Dominant features in the present Earth's reflection spectrum are shown in 
Figure~\ref{fig3}, 
where in this case the panels give the reflected intensity
normalized to the incident solar intensity, smoothed from the original
high-resolution calculation to a plotted resolution of 100.  
The upper left panel is a composite of the 5 contributing species H$_2$O, 
O$_2$, O$_3$, CH$_4$, and CO$_2$, for present abundances.  
Note that the spectral range goes 
from the near infrared (2000 cm$^{-1} = 5$ $\mu$m)  
to the near ultraviolet (33000 cm$^{-1} = 0.30$ $\mu$m),
with a continuum normalized everywhere to unity.  We have ignored the 
thermal emission contribution at long wavelengths as well as the Rayleigh and 
dust scattering components which will show up mostly at short wavelengths.

The H$_2$O panel, for the present abundance of water, shows a 
series of absorption bands
spanning the middle part of the visible spectrum and increasing in strength toward
the near infrared.   The strengths of these vibrational bands are essentially 
independent of temperature, but will increase in proportion to the abundance of 
water and the square root of air pressure, however since the lines are relatively
saturated, the average band depth will only increase as the square root of band
strength.  The net result is that these bands should be good indicators of the
presence of water over a large dynamic range of conditions, although this same
property makes them less useful as quantitative indicators of water mixing ratio,
unless we also have independent knowledge of temperature and pressure.

The strongest O$_2$ band is the Fraunhofer A-band at 0.76 $\mu$m (13000 cm$^{-1}$).  
This band too is saturated, and will still be relatively strong for significantly
smaller mixing ratios than the present Earth's.  It will therefore be an excellent 
indicator of the presence of oxygen (see Chapter VI.2).

The O$_3$ molecule has two broad features of note, the extremely strong Huggins band
which produces the ultraviolet absorption shown here at about 0.33 $\mu$m 
(30000 cm$^{-1}$) and shorter, and the Chappuis band which shows up as a broad 
triangular dip in the middle of the visible spectrum from about 0.45 to 0.74 $\mu$m 
(22000--13000 cm$^{-1}$).  
Ozone in the stratosphere is produced from O$_2$ molecules, and its abundance is a
non-linear function of the O$_2$ abundance, such that even a small amount of O$_2$ can
produce a relatively large amount of O$_3$ [{\it Kasting and Donahue}, 1980];
however in absolute terms the modeled column abundance of O$_2$ is
nevertheless large compared to that of O$_3$ by a factor of 20,000 to
500,000.  

Methane at present terrestrial abundance (1.65 ppmv) has no significant visible 
absorption features, but at high abundance (0.1\%) it has strong visible bands at
0.9 and 1.0 $\mu$m (11000 and 10000 cm$^{-1}$).

Carbon dioxide has negligible visible features at present abundance, but
in a high-CO$_2$ atmosphere (10\%) it has a significant band at 1.2 $\mu$m 
(8000 cm$^{-1}$) and even stronger ones at longer wavelengths.

\section{PLANETARY SPECTRA: NEOPROTEROZOIC EARTH} 

The present status of our secure knowledge of paleo-Earth atmospheres is easy 
to review, because little of our knowledge is secure, although in recent years 
the situation has been improving.  In overview, the evolution of CO$_2$ has 
long been believed to have decreased from a high level of roughly 1 bar at 
about 4.5 Ga (where Ga represents $1\times10^9$ years ago) to 0.00035 bar at 
present, however recent evidence suggests that 
major oscillations occured around 0.5-0.8 Ga, and perhaps at other glaciations, 
The abundance of O$_2$ is believed to have been less than 0.001 bar until
about 2 Ga when it rapidly began to increase toward its 0.2 bar present
value, and the time of transition is roughly coincident with the beginning of 
abundant phytoplankton on Earth.  

The equilibrium temperature of the Earth with its present albedo, present
solar flux, and no greenhouse gases is about 246 K, below the freezing point 
of water; this corresponds to the physical temperature at a level in the 
stratosphere at which the Earth effectively radiates.  The Earth is rescued
from freezing by the greenhouse effect, which can be envisioned as being 
driven at temperatures below the freezing point of water by CO$_2$
and aided at higher temperatures by evaporated water vapor.  The combination
of present levels of CO$_2$ and H$_2$O is sufficient to warm the surface to 
$290$ K, about the current average surface temperature, which is of
course sufficient for liquid water. 

The early Earth was illuminated by a weaker Sun, about 0.71 times the present
luminosity,  which without CO$_2$ would have resulted in an even cooler Earth, 
about $246(0.71)^{1/4} = 226$ K.  The argument for a large amount of early CO$_2$
is simply that we believe that liquid water, with only intermittent
glaciation, was present during most of Earth's past [{\it cf. e.g., Lunine} 1999,
Sec. 11.10 and 19.4], and this requires a large greenhouse effect, equivalent 
to about 0.2 bar or more CO$_2$.  Also, since there is the equivalent of about 
50 bar of CO$_2$ deposited in crustal rocks, the pressure may have been higher.

The level of CO$_2$ probably did not fall steadily since 4.5 Ga. 
We know that periods of major glaciation did occur, and that almost certainly
these could not have begun unless the CO$_2$ abundance had first dropped
significantly.  Recently
there has been a major advance in this area with the clear-cut
identification of 3 major glaciation cycles alternating with warm tropical
conditions at about 0.5--0.8 Ga [{\it Hoffman et al.}, 1998].
This period is roughly the end of the Proterozoic era (from 4.5 to 0.7 Ga),
and the start of the Phanerozoic era (from 0.7 to 0.0 Ga), so the sudden
glaciations are said to have produced a Neoproterozoic ``snowball'' Earth.
 The CO$_2$ abundances probably oscillated between roughly 0.2 bar
and 0.0001 bar.  The evidence for oscillation is unambiguously seen in rock 
layers in Namibia: repeated pairings of rounded boulders (from glaciers) 
topped by thick layers of carbonate rock (precipitated from seawater).

A major mystery is why these oscillations have not occurred over most of Earth's
history, or more pointedly, why the oscillations started when they did and
stopped after a few cycles.   It is possible that the other major glaciations 
were accompanied by such oscillations, 
but this has not yet been established, or even discussed, to our knowledge.

One might expect that the icehouse state would be triggered by a sudden
drop in CO$_2$, caused perhaps by a rapid uptake of CO$_2$ in the ocean, for some reason,
and that the hothouse state would be triggered by the accumulation of volcanic
CO$_2$, a direct result of plate tectonics, which at present rates would take only a
geologically short interval of about 10 million years to reach greenhouse levels.
 However the details of the oscillations are as yet not well understood.

\begin{figure}[tb]
\begin{center} 
\leavevmode 
\epsfxsize=3.25in
\epsffile[42 215 544 650]{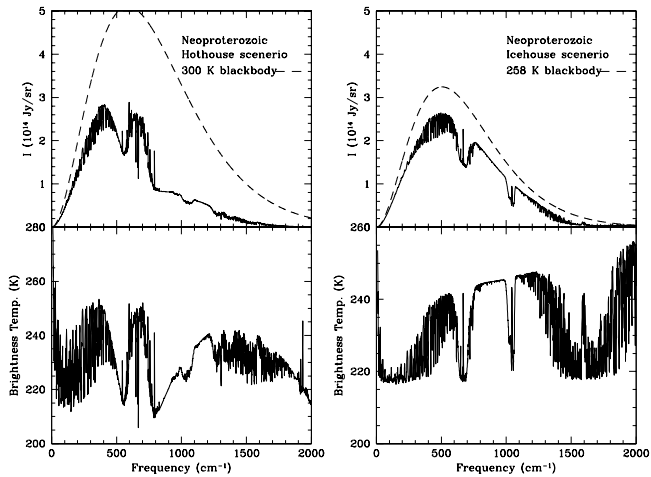}
\end{center} 
\caption{A calculated thermal infrared emission spectrum for a Neoproterozoic
hothouse condition (left), and icehouse (right).  Note the huge difference between
these;  for example, the 15 $\mu$m CO$_2$ feature is an apparent emission
feature in the hothouse spectrum, but it is an apparent absorption feature 
in the icehouse spectrum.
The apparent emission feature is a result of a warm inversion layer in the 
stratosphere (due to O$_3$ heating) combined with a CO$_2$ mixing ratio which 
gives an optical depth of about unity at the top of this layer for wavelengths 
in the core of the CO$_2$ band.}
\label{fig5} 
\end{figure}

The implication of the Neoproterozoic oscillations for extrasolar planet searching
is clear:  we might encounter such a condition on an extrasolar planet, and we must
be prepared to anticipate its spectral signature.  We show in 
Figure~\ref{fig5} 
our preliminary calculation of the infrared spectra of a Neoproterozoic 
hothouse and icehouse Earth.  To simulate the icehouse state, we use the present
Arctic temperature profile, and O$_2$ at 1\% PAL (present atmospheric level);  
the O$_3$, H$_2$O, N$_2$O, and CH$_4$ are modified 
[{\it Kasting}, 1980] for a 1\% O$_2$ level, and the CO$_2$ is set to 100 ppm; 
about 1 bar of N$_2$ is also assumed to be present.  To simulate the hothouse 
state, we used a present tropical temperature profile, O$_2$ at 10\% PAL, 
and corresponding O$_3$, H$_2$O, N$_2$O, and CH$_4$; also CO$_2$ is set to 
120,000 ppm.  

A substantial literature is building on the snowball Earth concept, for example
climate simulations [{\it Jenkins and Smith}, 1999; {\it Hyde et al.}, 2000] 
and realizations that life would not be totally eliminated by ice cover 
[{\it Runnegar}, 2000].

\section{PLANETARY SPECTRA: METHANE}

Besides CO$_2$, methane (CH$_4$) could strongly influence the surface temperature 
of a planet.  Methane bursts may have punctuated the Earth's history.
For example, carbon and oxygen isotope fractionation in marine carbonate deposits  
strongly imply that about 350 times the present level of CH$_4$ was injected into 
the Earth's atmosphere, during a span of less than 1000 years, 
about 0.055 Ga [{\it Bains et al.}, 1999].  The source of the CH$_4$ may well 
have been methane hydrate, which is found on the sea floor in abundance even 
today. That this time of injection coincided closely with a period of warming 
(the Late Paleocene thermal maximum) strongly suggests a cause and effect 
relationship between these phenomena, but it is not clear which is cause and 
which is effect.    

Further evidence for a methane burst has been found at another epoch, 0.183 Ga
[{\it Hesselbo et al.}, 2000].
This event is associated with high surface
temperatures and significant mass extinction.  
Yet other such events at 0.090 and 0.120 Ga have been similarly reported 
[{\it Kerr}, 2000].  
Interestingly, each of these events, now 5 in total, occurs at about the same
time that a large volume of volcanic outflow occured, and a corresponding mass
extinction occured, suggesting that perhaps the volcanoes triggered the methane
and this in turn triggered further warming which resulted in large-scale loss of
species.  Large amounts of CH$_4$, such as in these bursts, will produce
measurable absorption features in both the infrared emission and
near-infrared absorption spectra of the Earth.   

\section{PLANETARY SPECTRA: PRESENT VENUS AND MARS} 

We have calculated infrared spectra of Venus and Mars, using current models
of atmospheric abundances and temperature profiles, with
the added factor that the opacities of air-borne dust and aerosol droplets can
be included in order to reproduce observed silicate and ice features.
Clearly we have much less information on the geologic history of these planets
than we do for Earth, but spectral evidence of isotopic fractionation on both
planets has led to informed speculation about the past abundance of water, for
example.  

\begin{figure}[tb]
\begin{center} 
\leavevmode 
\epsfxsize=3.25in
\epsffile[42 215 544 650]{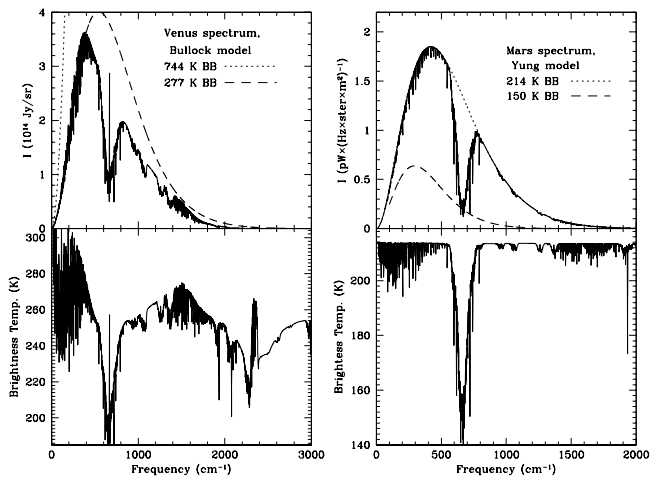}
\end{center} 
\caption{Calculated spectra for Venus and Mars. 
Both spectra are dominated by CO$_2$.  
No ice or dust features are included here, although these features are seen 
occasionally on Mars, and we have modeled them successfully. }
\label{fig7} 
\end{figure}

We show in the right panel of Figure~\ref{fig7}
our calculated Mars spectrum using a recent model atmosphere 
[{\it Y. Yung}, 1999, private communication].  
The CO$_2$ and H$_2$O features in this spectrum are a good match to many
observed Mars spectra.  We have repeated these calculations with the inclusion 
of thin ice clouds and silicate dust clouds, and these provide a reasonably
good simulation of certain other observed Mars spectra.
The calculations for Venus are shown in the left panel of Figure~\ref{fig7},  
based on the model atmosphere of {\it Bullock and Grinspoon} [2001].
This spectrum is dominated by CO$_2$, and does not include any aerosols, 
so the fact that the very bright thermal emission from the ground is not seen
at the top of the atmosphere is entirely due to the opacity of the wings of
the strong CO$_2$ bands.

\section{CAN WE DETECT LIFE?} 

The big question, can we detect life on an extrasolar planet, was addressed 
in a general sense by {\it Lovelock}, [1965]
who advocated searching for signs of chemical non-equilibrium, 
such as the simultaneous presence of reducing and oxidizing gases.
In principle, if one knew the abundances of all species present in an
atmosphere, and the boundary conditions (incident spectrum from star,
surface composition, rotation rate, etc.), then one could calculate the
state of photochemical and dynamical equilibrium, and its likely
fluctuations, and compare this with the observed state, to see if there 
are significant differences which might be interpreted as signs of life,
a term not included in the equilibrium calculation
[{\it cf. Nisbet and Sleep}, 2001; {\it Brack}, 1998].
However our experience with the atmosphere of the present Earth tells us 
that the observed state is frequently not predicted by theory,
so this approach may not be foolproof 


From a spectroscopic point of view, it is natural to think of searching for
life in terms of measuring the abundances of atmospheric constituents, and 
comparing these abundances with the results of calculations of 
thermochemical and photochemical equilibrium.  In particular, the 
simultaneous presence of significant amounts of oxidized and reduced
species, such as H$_2$O, CO$_2$, N$_2$O, and CH$_4$, or the presence of 
large amounts of O$_2$ or O$_3$, would both be indicators of life. 
Indeed both conditions are found on Earth, where there is no known means
of producing anything approaching a 21\% O$_2$ atmosphere except by
photosynthesis, and the main sources of N$_2$O and CH$_4$ are biological.
Earth's O$_3$ is produced in relatively large amounts in the stratosphere
by photolysis and recombination of O$_2$, making it a good indicator of
the presence of O$_2$.
As signs of life in themselves, H$_2$O and CO$_2$ are secondary in importance, because although they are 
raw materials for life, they are not unambiguous indicators of its presence.
Farther down the chain, CH$_4$ is a life product on Earth, but elsewhere 
it is also a ubiquitous 
primordial species, and it is difficult to detect
spectroscopically unless it is quite abundant.  
Likewise, N$_2$O is interesting because it is produced in abundance
by life, and only in trace amounts by natural processes, but it can only
be detected in a region which is strongly overlapped by CH$_4$ and H$_2$O, and
so is an unlikely prime target. 

As counterexamples, we know that small amounts of O$_2$ and O$_3$ are 
readily produced by non-biological means.   
For example, a trace amount of O$_2$ is seen on Venus, consistent with 
photochemical production from CO$_2$;  on Mars both O$_2$ and O$_3$ are 
measured, but the amounts are in agreement with calculations of photochemical
production from CO$_2$ and loss due to CO$_2$ and H$_2$O and products 
[{\it e.g., Yung and DeMore}, 1999].
 In two other examples, Europa has about $10^{-11}$ bar of O$_2$ in its
atmosphere (e.g., Chapter III.3), and Ganymede may have O$_2$ trapped in 
surface ice, but in both cases the observed signatures are consistent 
with energetic particle bombardment of a water-ice surface, and therefore 
they do not require a biological source. 
 
An intriguing theory of the origin of life on Earth [{\it Wachtershauser}, 2000]
has found experimental backing in new laboratory experiments 
[{\it Cody et al.}, 2000] which show that iron sulfide at elevated pressure and
temperature can facilitate the natural generation of organometallic compounds by
an autocatalytic process, in which the products of reaction catalyze the next
cycle of reaction, making the process likely to be self-sustaining.
This work is of interest because iron and sulfur atoms are at the center of many
enzymes, and the process provides a simple explanation of their presence in
living cells.  The work is also of interest because it suggests that the origin
of life on Earth may have been deep underground, where conditions are right for
the reactions to take place, and out of the possibly less-friendly conditions on
the surface of the early Earth.  

The Neoproterozoic oscillations discussed above may also be critically 
important to the development of life on Earth, because, as the eponymic
oscillations suggest, this era coincided with the explosive radiation of life
forms on Earth, from simple algae types to the dramatically different types of
life forms we know today.  It is thought that the huge stresses, due to
oscillations of temperature, nutrients, and available sunlight, imposed on simple
life forms forced them to develop new adaptations in order to survive in their
new environments, in what may have been the greatest Darwinian experiment ever. 
Thus, if we find extrasolar planets in the throes of such oscillations, we may expect
that a similar life-radiation may be occuring on that planet, although we will
have to wait another several tens or hundreds of million years before we can see
how the experiment turns out.

\acknowledgments
This work at SAO was supported by NASA via contract JPL 1201749, 
and via the TPF program through a contract to Ball Aerospace \& 
Technologies Corp. 

  

\newpage

\vfill \eject

\vfill \eject

\vfill \eject

\end{article}
 
\end{document}